\documentclass[journal,comsoc]{IEEEtran}
\usepackage[T1]{fontenc}

	 \usepackage{graphicx}
	\usepackage{epstopdf}

	\usepackage{amsmath}

	\usepackage[cmintegrals]{newtxmath}
	
	\usepackage{bm}

	\usepackage{algorithm}
	\usepackage{algorithmic}

	\usepackage{array}
	
	\usepackage{url}

	\usepackage{multirow} 
	\usepackage{bigdelim}
	\usepackage{leftidx}  
	\usepackage{multicol} 
	\usepackage{bibentry}
	\usepackage[]{footmisc}	
	\usepackage{romannum}
	\usepackage{geometry}
\usepackage[switch]{lineno} 

\usepackage[export]{adjustbox}  

\makeatletter
\def\blfootnote{\xdef\@thefnmark{}\@footnotetext}
\makeatother

\begin{document}

\title{Differential Transmission Schemes for Generalized Spatial Modulation }

\author{\IEEEauthorblockN{Deepak Jose,~\IEEEmembership{Student~Member,~IEEE,} Sameer S. M.,~\IEEEmembership{Senior~Member,~IEEE}} \\
\IEEEauthorblockA{National Institute of Technology Calicut, India}}

\maketitle
\blfootnote{Copyright (c) 2015 IEEE. Personal use of this material is permitted. However, permission to use this material for any other purposes must be obtained from the IEEE by sending a request to pubs-permissions@ieee.org. Citation information: DOI 10.1109/TVT.2021.3118457, IEEE
Transactions on Vehicular Technology. Authors would like to thank the Department of Science \& Technology, Government of India for supporting this work under the FIST scheme No. SR/FST/ET-I/2017/68}
\begin{abstract}
Differential modulation schemes are very relevant in receivers having power and processing limitations, as these schemes dispense with the need for knowledge of channel coefficients for symbol detection. Spatial modulation (SM) is a scheme used in multi-antenna transmission scenarios where the data is transmitted in the amplitude, phase and spatial domains through selected antennas. In the coherent domain, generalized SM (GSM) employs multiple antennas in combination during every time slot to enhance the spectral efficiency (SE). In this paper, we propose two differential schemes which activate two or more antennas at a time to transmit the modulated symbol. These schemes achieve higher SE using a lesser number of antennas and lower order modulation schemes instead of the higher number of antennas required for conventional SM schemes based on differential modulation. Simulation studies reveal that the proposed schemes have better bit error rate performance than traditional differential SM schemes. We also derive the analytical union bound for the proposed schemes and is satisfied from medium to high  signal-to-noise ratio (SNR) ranges.
\end{abstract}
\pagenumbering{arabic}
\begin{IEEEkeywords}
Spatial modulation (SM), generalized spatial modulation (GSM), differential spatial modulation (DSM), multiple input multiple output (MIMO), average bit error probability (ABEP)
\end{IEEEkeywords}
	
\nolinenumbers  
	
\section{Introduction}
Information and communication technology (ICT) sector contributes to  $2 \%$ of the world's carbon emissions \cite{sm-survey1}. The ever-growing number of base stations (BSs) numbering more than $5$ million  around the globe \cite{greensurvey1}, adds up to the energy consumption of the cellular operators. The radio network side alone consumes $80 \%$ of the operator's total power requirement and a BS consumes an average of $25$ MWh per year \cite{greensurvey2}. In existing communication  systems, a dedicated radio frequency (RF) chain is available at the transmitter to drive every antenna. The RF chains which contain different types of power amplifiers (PAs) consume up to $65 \%$ of the transmitter circuitry power \cite{deliotte2010}. Spatial modulation (SM) can help in reducing the need for multiple RF chains at the transmitter hardware by sharing one RF chain among all the transmit antennas. Also, it is proposed that base stations (BSs) employing SM can have up to  $67 \%$ energy efficiency \cite{sm_energy_eff} compared to BSs using conventional transmission schemes such as space-time block codes (STBC), multiple input multiple output (MIMO) and multiple input and single output (MISO) schemes, as SM avoids inter-antenna synchronization and the need for multiple RF chains to power the transmitter antennas. 
\par
In SM, a single antenna is activated at each time instant based on the information bits and an M-ary quadrature amplitude modulation (MQAM) or M-ary phase shift keying (MPSK) symbol conveys the other part of the information bits through the active antenna, thereby conveying the information in the spatial, phase and amplitude domains \cite{sm_jeg}. Generalized spatial modulation (GSM) \cite{gsm1}, \cite{gsm_2012} is a modified type of SM that involves activating two or more antennas at the same time instant to transmit the modulated symbols. The flexible architecture of GSM enables the use of an arbitrary number of transmit antennas, unlike SM that restricts the antenna number to the power of two (eg:- $2,4,8,16, \ldots$). The enhanced SM (ESM) in \cite{esm_2015} uses a primary and secondary constellation to transmit the information bits when the number of active antennas is varied at the transmitter. The variable active antenna-based  scheme of \cite{var_Nu_gsm_2017} also improves the error performance in comparison to GSM. The variable antenna schemes of  \cite{esm_2015, var_Nu_gsm_2017} increases the ambiguity while detecting the TAC, and it is solved using the block-based SM (BSM) proposed in \cite{new_block_sm_2018} and \cite{block_sm_constl_2020}. This block-based approach translates to increased detection complexity for higher order modulation.
 In another work, the information bits are jointly mapped to a varying number of active antennas using more than one RF chain \cite{joint_var_ant_sm_2020}, unlike the independent mapping in the previously discussed schemes where the bits are mapped separately to an antenna combination and a modulation symbol. Also, the joint mapping of the bits to the active antennas and constellation symbols helps in improving the error performance compared to GSM.  
The scheme precoded-SM (PSM) \cite{precod_sm_NOMA_2021} improves the spectral efficiency (SE) by using the channel state information (CSI) at the transmitter. Quadrature SM (QSM) \cite{qsm_2015}, is the first SM based scheme to employ two separate active antennas to transmit the in-phase (I) and quadrature-phase (Q) of the same modulation symbol and it has a few variants as well,  such as complex-QSM \cite{c_qsm_2017}, double-SM \cite{double_sm_2016}, improved-QSM \cite{I_qsm_2017} and signed-QSM \cite{s_qsm_2020} which improves the overall spectral efficiency in comparison to SM, though some of these schemes achieve an improvement at the cost of an increase in detection complexity. 
\par
Conventional differential SM (DSM) \cite{udsm}, \cite{1stdsm} is a promising alternative to coherent SMs, since it does not require CSI at the detection stage. DSM avoids some overheads also, as it does not require training symbols and computations for channel estimation (CE) and it is free from associated  errors. However, like most of the traditional differential schemes, DSM also suffers from a $3$ dB penalty in error performance as compared to SM. Also, the unitary property of the DSM symbols and the encoding procedure creates a large symbol map for any slight increase in the modulation order, thus making the detection process computationally prohibitive. To reduce the detection complexity, low complexity schemes based on hard-limiting maximum-likelihood (HL-ML) \cite{lowdsm1} and compressive sensing \cite{cs_dsm} are implemented.
 
\par
Later on, amplitude phase shift keying aided DSM (APSK-DSM) schemes are introduced in \cite{dsm_apsk1} and \cite{dsm_4_ring_apsk} to improve the throughput, where the improvement is at the cost of an exponentially increasing detection complexity for an incremental change in the modulation order. Though a low complexity detector is developed in \cite{cs_apsk_dsm_iet}, the transmission scheme, in general, suffers from its inherent error propagation issue as the previously detected symbols are used for the detection process. Moreover, a general framework to scale up the amplitude domain constellation to higher order does not exist for these schemes. An alternative approach to modulate amplitude information in the differential domain is introduced through differential quadrature SM (DQSM) \cite{dqsm_2017}. Here the unitary properties are  followed only for a limited modulation order of  $M=2, 4$. This limitation can be overcome only by increasing the transmit antenna number, thus restricting this scheme to base stations.
\par
A parallel approach to develop	a differential scheme for SM is found in generalized differential scheme for SM (GD-SM), where the unitary properties of the symbols are not necessary for encoding \cite{gdsm}. This difference in encoding allows MQAM symbols to be used in the transmission, thereby making GD-SM a spectrally efficient scheme. In contrast to the DSM and APSK-DSM techniques, the absence of unitary property allows the GD-SM symbol to be transmitted in a single slot, which also helps in maintaining the symbol map size to a reasonable level while increasing the modulation order. This significant advantage over the existing differential schemes \cite{1stdsm}, \cite{dsm_apsk1} and \cite{dsm_4_ring_apsk} reduces the detection complexity of the optimal detector for GD-SM. In order to further reduce the complexity, authors in \cite{low_gdsm_spcom} have suggested a low complexity detector for the scheme. Moreover, the conventional schemes such as SM, along with  GD-SM, achieve a higher SE only by increasing the transmit antennas as a power of two, thereby naturally increasing the hardware cost and form factor of the transmitter.
\par
Most of the differential SM schemes have their symbols represented as square matrices in the system model, and this requires multiple time intervals to transmit a symbol. In order to conserve the channel usage and to exploit the multiple antennas at the transmitter, rectangular DSM (RDSM) is proposed \cite{rect_dsm_2017}, where the square matrices of conventional DSMs are transformed to rectangular matrices. Another variant of RDSM \cite{rdsm_near_coh_2020} employs an adaptive forgetting factor to reduce the impact of error propagation which is common to all RDSMs. Hence, reset symbols are employed at periodic intervals to minimize the error propagation similar to APSK-DSMs \cite{dsm_apsk1}, \cite{dsm_4_ring_apsk}. Moreover, these schemes require a well optimized forgetting factor which is to be designed for the specific channel environment, and it demands extra processing overhead at the receiver.
\par
Nearly all the differential schemes discussed so far are unable to overcome the well known SNR penalty in the range of  $2$ to $3$ dB with  their coherent counterparts for the same bit error rate (BER). We propose two differential encoding schemes employing multiple active antennas to transmit the symbols with the motivation to increase  the throughput and narrow down the error performance penalty compared to the coherent schemes, as well as to improve the SE compared to the GD-SM scheme. Unlike the GD-SM and SM \cite{sm_jeg} schemes that are restricted to the use of transmit antennas only as of the power of two, the proposed schemes are capable of employing an arbitrary number of transmit antennas, thereby supporting devices having limited form factor. In contrast to the conventional differential encoding techniques which use consecutive symbols for detection, the proposed schemes use the reference symbols transmitted at the start of a frame to encode all other normal symbols of the frame, thus avoiding the error propagation which is common in existing differential schemes such as APSK-DSMs, and the variants of RDSM. Even though the reference signals are a slight overhead, it is negligible when compared to GD-SM, APSK-DSMs and RDSMs. Though inter antenna synchronization is also required at the transmitter for the proposed schemes, unlike the single active antenna schemes, the significant improvement in SE and increased throughput justifies this extra transmitter processing. Also, during unfavourable or noisy channel conditions where the  modulation order is required to be fixed, we show that the proposed schemes sometimes require either minimal or no increase in the number of transmit antennas to support a given SE, in contrast to the conventional differential schemes such as DSM, APSK-DSMs and GD-SM.  Further, we derive an upper bound on the bit error performance for an arbitrary number of transmit and receive antennas using an MQAM or MPSK modulation. The bound obtained is very tight for a wide range of SNRs. 
\par
The contributions of this paper are summarized as follows:-
\begin{itemize}
\item We propose two differential schemes which employ multiple active antennas to transmit the modulated symbols, unlike the single antenna schemes in the existing literature. The first one among the proposed methods, named differential-GSM (D-GSM), uses only one RF chain and the second one, its high throughput version, termed as differential-multi generalized spatial modulation (D-MGSM) employs more than one RF chain to drive the transmit antennas. It is found that the BER performances of the proposed schemes are not only better than GD-SM but also close to some of the coherent schemes employing multiple active antennas. They also dispense with the need for CSI at the receiver (CSIR) thereby improving the throughput.\\ 
\item The detection complexity of the proposed schemes is analyzed and compared with that of the existing detectors in the differential and coherent domains. \\
\item An upper bound on the error performance of the proposed scheme is derived using the moment generating function approach, and the bounds are tighter from the mid-SNR region onwards. The derived bound is valid for an arbitrary number of transmit and receive antennas. \\
\item A power allocation method is adopted to drive the multiple active antennas, in contrast to the equal power allocation in the single antenna schemes of DSMs and APSK-DSMs.\\

\end{itemize}
\par
 The remainder of this paper is organized as follows: Section \ref{sec:existing_sm}, describes the system model for DSM, GD-SM and GSM. The system model for the proposed schemes along with the power allocation strategy is described in Section  \ref{sec:prop_sm_model}, which also describes the optimal detectors for both the schemes. The analytical union bound on the BER is derived in Section \ref{sec:ABEP}. Computational complexities of the proposed detection schemes are analyzed in Section \ref{sec:complexity}. Results of extensive simulation studies are presented in Section \ref{sec:simulation} followed by the conclusion in Section \ref{sec:conclusion}.

	\par
\textit{Notations:}~  ${\lVert . \rVert}_{p}$, ${(.)}^T$, ${(.)}^H$, ${(.)}^\dag$ and $|.|$ denote the $\ell_{p}$ norm, transpose, Hermitian, pseudo-inverse and absolute value operations on a vector or matrix. $\lfloor . \rfloor $ represents the floor operation and $\dbinom{.}{.}$ represents the binomial coefficient. Boldface upper and lower case letters represent matrices and vectors. The symbol $!$ stands for the  factorial operation when it succeeds a number or variable, and $!!$ denotes semifactorial. The number of  transmitter and receiver antennas are denoted by $M_t$ and $M_r$. Any real or complex numbers are denoted by $\mathbb{R} $ and $\mathbb{C} $, respectively.

\section{Conventional SM schemes} \label{sec:existing_sm}
In this section, we briefly review the first DSM scheme in the literature to differentiate the encoding scheme of our proposals. Then the GD-SM scheme is discussed to demonstrate its limited symbol map size, thus simplifying the decoding. Finally, the coherent counterpart of the proposed transmission technique is also described for better understanding.

\subsection{Conventional DSM}
Using $M_{t}$  transmitter antennas,  the differential encoding of DSM \cite{1stdsm} is given by  
\begin{equation}			
			 {\bf S}_{k} = {\bf S}_{k - 1} {\bf X}_{k}	
\end{equation}
where ${\bf S}_{k} \in {\mathbb{C}}^{M_t \times M_t}$ is the transmitted symbol and ${\bf X}_k \in {\mathbb{C}}^{M_t \times M_t}$ is the original DSM symbol whose structure is $ {\bf X}_{k} = \begin{bmatrix} 0 & x_{12}  \\x_{21} & 0  \\   \end{bmatrix}$, for $M_t =2$ where $x_{21}$ and $x_{12}$ are the MPSK symbols transmitted during two consecutive time intervals. The row number corresponding to every non-zero element in the column of ${\bf X}_k$ points to the transmit antenna that is active in  the specific time interval. The receiver model is described as
\begin{equation}	
		 {\bf Y}_{k} =   {\bf H}{\bf S}_{k} +  {\bf N}_{k}	
\end{equation}
where  ${\bf Y}_{k}$ $\in {\mathbb{C}}^{M_r \times M_t}$, is the received symbol in the $k^{th}$ block, ${\bf H} \in {\mathbb{C}}^{M_r \times M_t}$, is the channel matrix and  ${\bf N}_{k} \in {\mathbb{C}}^{M_r \times M_t}$, is the noise matrix. 

\subsection{GD-SM }
In this scheme \cite{gdsm}, the symbols are transmitted in frames containing $L = M_t + K$ symbols each. The reference symbol $s^r = 1$ is transmitted for each of the antennas at the start of the frame without repetition, totalling to $M_t$ symbols, followed by the $K$ normal symbols carrying the information bits. Thus, one reference symbol received during the $i^{th}$ $(i \in [1,M_t])$ time instant is denoted by ${\bf y}^r_{i} \in {\mathbb{C}}^{M_r \times 1}$ and  is defined as, 
\begin{equation}
{\bf y}^r_{i} = {\bf h}^r_{i}s^r + {\bf n}^r_{i}
\label{eq:gd_sm_ref_rx_sg}
\end{equation}
where  ${\bf h}^r_{i} \in {\mathbb{C}}^{M_r \times 1}$ is the channel matrix during the $i^{th}$ time instant and  ${\bf n}^r_{i} \in {\mathbb{C}}^{M_r \times 1}$ is the noise vector. 
The individual entries of ${\bf h}^r_{i}$ and ${\bf n}^r_{i} $ are independent and identically distributed (i.i.d) as  $\mathcal{CN}(0,\, 1)$ and $\mathcal{CN}(0,\sigma^{2}_r)$ respectively.
The normal symbol $x$, carries the information as an MPSK or  MQAM symbol, and hence the GD-SM  symbol is denoted as ${\bf x}^{n} = {\lbrace 0 \ldots x \ldots 0 \rbrace}^T$ and has only one non-zero symbol located at the transmit antenna number that is activated. The differential encoding for the normal symbol is given by 	${\bf s}^n = s^r {\bf x}^{n}$, and hence the received signal ${\bf y}^{n} \in {\mathbb{C}}^{M_r \times 1} $ is denoted as   
\begin{equation}
	{\bf y}^{n} = {\bf H}^n {\bf s}^n + {\bf n}^{n}
	\label{eq:gd_sm_normal_rx_sg}
\end{equation}
where ${\bf H}^{n} \in {\mathbb{C}}^{M_r \times M_t}$ is the channel fading matrix and ${\bf n}^{n} \in {\mathbb{C}}^{M_r \times 1} $ is the additive white Gaussian noise (AWGN) vector during the the normal symbol blocks.

\subsection{Conventional GSM} \label{sec:sm_model_gsm}
The first scheme for SM, which activates multiple antennas at a time is proposed in \cite{gsm1}, where the communication system has  $M_t$ transmitter antennas, of which  $M_u$ number of antennas are  activated at a time, by sharing a single RF chain. During each symbol transmission, these $M_u$ antennas shall send identical MQAM or MPSK symbol. Whereas in \cite{gsm_2012}, $M_u$ number of RF chains are utilized to activate that many antennas, to transmit unique modulation symbols belonging to the modulation order $M$. Thus the transmitted symbol is represented in general form as ${\bf s} = {\lbrace 0 \ldots x_1  \ldots 0 \, x_{M_u} \ldots  \rbrace}^T$,  where the non-zero entries correspond to the symbols transmitted in the $M_u$ active antennas.

The received symbol ${\bf y} \in {\mathbb{C}}^{M_r \times 1}$ is written in general form as,
\begin{align}
	\begin{split}
	{\bf y} &= {\bf H}{\bf s} + {\bf n} \\	
 			&= \sum_{i=1}^{M_u} \boldsymbol{h}_{l_i}x_i + {\bf n}	
		\label{eq:gsm_model}
	\end{split}
\end{align} 
where  $ \boldsymbol{h}_{l_i}$ is the $l_i^{th}$ column of the fading channel matrix ${\bf H}$  and ${\bf n} \in {\mathbb{C}}^{M_r \times 1} $ is the AWGN vector whose entries  are i.i.d as $\mathcal{CN}(0,\sigma^{2})$.

\section{Proposed Transmission Schemes} \label{sec:prop_sm_model}
\subsection{System model}
We propose two schemes having multiple active antennas to transmit the information, unlike the single active antenna scenario of GD-SM, SM, DSM, APSK-DSM and DQSM. Out of the  $M_t$ transmit antennas, $M_u$ of them are activated at a time to create $M_c = \dbinom{ M_t}{M_u} $ combinations. However, we can use only the power of two antenna combinations totalling to $m_c = 2^{\lfloor  log_2{M_c} \rfloor} $ as the valid antenna combinations. Given below is an example of a transmitter with $M_t = 5$ antennas using $M_u = 2$ active antennas at a time to achieve the information mapping shown in  Table \ref{tab:g_dsm_tac_map}. 

\begin{table}[ht]	  
	 \centering 	
 	\caption{Mapping rule of information bits to transmit antenna for D-GSM and D-MGSM}
 	\hspace{-0.5cm} 	
		\begin{tabular}{l|llllllll}			
				\hline
				Information  & \multirow{2}{*}{$000$} & \multirow{2}{*}{$001$} & \multirow{2}{*}{$010$} & \multirow{2}{*}{$011$} & \multirow{2}{*}{$100$} & \multirow{2}{*}{$101$} & \multirow{2}{*}{$110$} & \multirow{2}{*}{$111$}  \\
				 bits & & & & & & & &  \\
				\hline		
				$\mathbb{L}$, Antenna & \multirow{2}{*}{$1,2$} & \multirow{2}{*}{$1,3$} & \multirow{2}{*}{$1,4$} & \multirow{2}{*}{$1,5$} & \multirow{2}{*}{$2,3$} & \multirow{2}{*}{$2,4$} & \multirow{2}{*}{$2,5$} & \multirow{2}{*}{$3,4$} \\
				combination & & & & & & & &  \\
				\hline		
		\end{tabular}		
  		\label{tab:g_dsm_tac_map}
 	\end{table}	

Both the proposed schemes have the same frame structure consisting of $M_t$ reference symbols followed by $K$ normal symbols carrying information. The schemes employ a reference symbol $s^r = a\, constant$, across each of the transmit antenna during the start of the frame. Thus the reference symbol block received at the start of the frame is given by
\begin{equation}
{\bf Y}^r = {\bf H}^{r}s^r + {\bf N}^r .
\label{eq:prop_ref_block_rx_sg}
\end{equation} 
where the  entries of the channel fading matrix ${\bf H}^r$  and the AWGN matrix  ${\bf N}^r \in {\mathbb{C}}^{M_r \times M_t}$, are i.i.d with distributions $\mathcal{CN}(0,\, 1)$, $\mathcal{CN}(0,\sigma^{2}_r)$ respectively.

\begin{table}[ht] 	 	
 	\caption{Characteristics of the proposed schemes, along with the conventional coherent and differential spatial modulation schemes}
 		\hspace{-0.5cm}
		\begin{tabular}{l|llll}
		\hline
\multirow{2}{*}{Scheme} & Pilot Overhead  & Error & \multirow{2}{*}{CSIR} & Symbol      \\	
    	   & (Reference Signals) & Propagation & & map size  \\
		\hline 
GSM\cite{gsm1, gsm_2012} & Long & No  & Yes & Small\\	
GD-SM\cite{gdsm}  &Long & No  & No  &  Small\\ 		
DSM \cite{1stdsm}  &No & No  & No   & Large\\ 
APSK-DSM &\multirow{2}{*}{Short} & \multirow{2}{*}{Yes}  & \multirow{2}{*}{No} &  \multirow{2}{*}{Large}\\
 \cite{dsm_apsk1, dsm_4_ring_apsk} & & & & \\
RDSM  &\multirow{2}{*}{Short} & \multirow{2}{*}{Yes}& \multirow{2}{*}{No} & \multirow{2}{*}{Medium} \\
\cite{rect_dsm_2017, rdsm_near_coh_2020} & & & & \\
Proposed   &\multirow{2}{*}{Short} & \multirow{2}{*}{No} & \multirow{2}{*}{No} & \multirow{2}{*}{Small}\\
Schemes   & & & & \\
		\hline
		\end{tabular}
  		\label{tab:charac_feat_all}
 	\end{table} 	

In Table \ref{tab:charac_feat_all}, the abstract level features of the proposed scheme is compared with the existing differential schemes  along with the coherent schemes in SM. It can be seen that, GD-SM alone has a few matching performance metrics with the proposed scheme. 

\subsection{Proposed  Differential-GSM (D-GSM) scheme} \label{sub:d_gsm}
This scheme requires only one RF chain, as it transmits the same modulation symbol across the active  antennas at the same time. By sharing the same RF chain among all the antennas, this scheme combines the advantage of a reduction in hardware similar to GD-SM, DSM and SM along with the increased SE guaranteed through the antenna combinations shown in  Table \ref{tab:g_dsm_tac_map}. Still, the active transmit antennas must be synchronized to avoid inter-symbol interference (ISI). Information is transmitted via the $K$ normal symbols. Symbol at the $k^{th}$ time slot where $k \in [1, \, K]$ is encoded as ${\bf x}^{(k)} = {\bf x}^{(k)}_{G} =  {\lbrace 0 \ldots x^{(k)}  \ldots \, 0 \, x^{(k)} \ldots 0 \rbrace}^T$, where the same symbols are positioned at the location of the $M_u$ active antennas. The differentially encoded vector for the normal symbol is given by 
\begin{equation}
	{\bf s}^n = s^r {\bf x}^{(k)}.
	\label{eq:prop_sg_encode}
\end{equation}
where  $s^r = a\, constant$. Consider the case of transmitting the bits $[1~1~0~0]$ using the D-GSM symbol using $M_t=5$ and $M_u=2$ antennas. The first three bits $[1~1~0]$ map to the TAC $(2,5)$ given in the Table \ref{tab:g_dsm_tac_map}, and the last bit '$0$', is modulated using BPSK, and is repeatedly transmitted over the above two active antennas such that the final symbol looks like ${\bf x}^{(k)} = [ 0~{1}~0~0~{1}]^{T}$. Here the non-zero entries correspond to the active antennas selected from the TAC mapping table.
The received signal ${\bf y}^n \in {\mathbb{C}}^{M_r \times 1} $ is denoted by  
 \begin{equation}
	\label{eq:d_gsm_normal_rx_sg}
	 {\bf y}^n = {\bf H}^n {\bf s}^n + {\bf n}^n           
    \end{equation}
where ${\bf H}^n \in {\mathbb{C}}^{M_r \times M_t}$ is the channel matrix and ${\bf n}^n \in {\mathbb{C}}^{M_r \times 1} $ is the AWGN vector, whose entries are i.i.d with distribution  $\mathcal{CN}(0, \, 1)$ and $\mathcal{CN}(0,\,\sigma^{2}_n)$ respectively.
The received signal can also be written in terms of the active antennas as
\begin{align}
\label{eq:d_gsm_normal_rx_sg_intermed_1}
	\begin{split}
{\bf y}^n &= {\bf h}^n_{l_1} s^r x^{(k)} + \ldots + {\bf h}^n_{l_{M_u}} s^r x^{(k)}  + {\bf n}^n \\
 &= \sum_{i=1}^{M_u} \boldsymbol{h}^n_{l_i}s^{r}x^{(k)} + {\bf n}^n	
	\end{split}
\end{align}
where ${\bf h}^n_{l_{i}}$'s are the column vectors of ${\bf H}^n$ corresponding to the active antennas alone  and the active transmit antenna combination is denoted by $\mathbb{L} = [l_1,l_2, \ldots, l_{M_u}]$. We assume that the channel is quasi-static during each frame. Thus the channel matrices during the reception of the reference and the normal data signals are approximately the same, i.e., ${\bf H}^r \approx {\bf H}^n $. Based on this assumption, we can rewrite the received normal data symbol ${\bf y}^n$ in terms of the received reference symbols ${\bf y}^r_{l_i}$ as,
\begin{equation}
\label{eq:d_gsm_normal_rx_sg}	
	{\bf y}^n     = \sum_{i=1}^{M_u} {\bf y}^r_{l_i} x^{(k)} - \sum_{i=1}^{M_u} {\bf n}^r_{i} x^{(k)} + {\bf n}^n.
\end{equation}

The ML based detector is thus defined as 
\begin{equation}
	[\hat{\mathbb{L}},  {\hat{x}}^{(k)}] =  \arg\min\limits_{\forall x^{(k)} \in\mathcal{G}, \forall {\mathcal{L}}} ||\boldsymbol{y}^n -  \sum_{i=1}^{M_u} \boldsymbol{y}^r_{l_i}x^{(k)} ||_{F}^{2} 
	\label{eq:d_gsm_ml_det}
\end{equation}
where a search is performed for finding the transmit antenna combination (TAC) $\hat{\mathbb{L}}$, from the TAC map given in $\mathcal{L}$  and the amplitude and phase modulated symbol $x^{(k)}$ is searched in the symbol map $\mathcal{G}$. The number of bits per symbol for this scheme  is  
$log_2{M} + \lfloor log_2{\dbinom{M_t}{M_u}} \rfloor$, which is also denoted as the number of bits per channel use (\textit{bpcu}). The term \textit{bpcu} is defined as the number of bits transmitted through the channel in a single instant of time, which also denotes the effective channel usage. 

\subsection{Proposed Differential Multi-GSM (D-MGSM) Scheme} \label{sub:d_mgsm}
D-MGSM scheme is a high throughput version, which employs $M_u$ RF chains to distinctly drive the active antennas. The information is encoded as ${\bf x}^{(k)} = {\bf x}^{(k)}_{MG} = {\lbrace 0 \ldots x_1^{(k)} \, x_i^{(k)}  \ldots \, 0 \, x_{M_u}^{(k)} \ldots 0 \rbrace}^T$, where the non-zero entries correspond to the antennas that are activated to transmit the modulated symbols $x_i^{(k)}$. 
 The differential encoding of the symbol follows (\ref{eq:prop_sg_encode}), and the received signal model is similar to (\ref{eq:d_gsm_normal_rx_sg}), where the repeating modulation symbols of D-GSM is replaced by distinct symbols in D-MGSM as,

\begin{align}
\label{eq:d_mgsm_normal_rx_sg_intermed_2}
	\begin{split}
	{\bf y}^n &= ({\bf y}^r_{l_1} - {\bf n}^r_{1}) x^{(k)}_1 + \ldots + ({\bf y}^r_{l_{M_u}} - {\bf n}^r_{{M_u}}) x^{(k)}_{M_u}  + {\bf n}^n \\
              &= \sum_{i=1}^{M_u} {\bf y}^r_{l_i} x^{(k)}_i - \sum_{i=1}^{M_u} {\bf n}^r_{i} x^{(k)}_i + {\bf n}^n
	\end{split}
\end{align}

\begin{figure}[h]	
		\begin{center}		
		 \includegraphics[width=9cm, height=8.2cm]{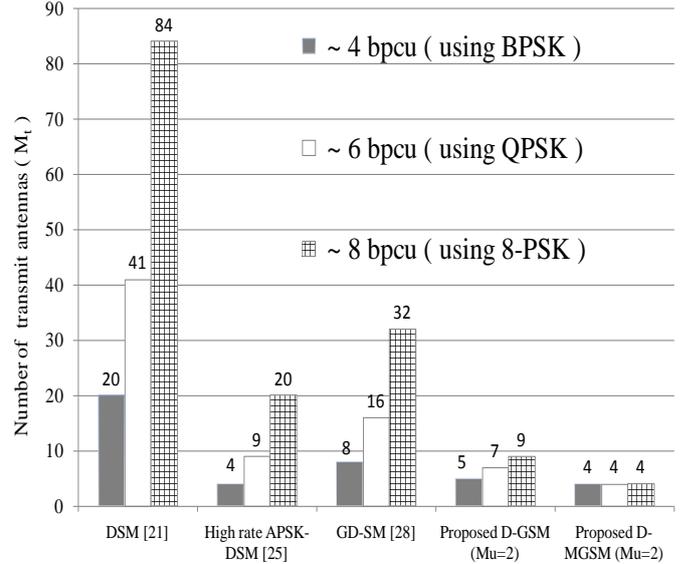} 
    \caption{Comparison of the number of transmit antennas required to achieve a given SE after fixing the modulation order  }
		  \label{fig:mt_versus_bpcu}
		\end{center} 
\end{figure}

Thus using the ML principle, the spatial and modulated symbols are jointly decoded using the equations (\ref{eq:prop_ref_block_rx_sg}), (\ref{eq:prop_sg_encode}) and (\ref{eq:d_mgsm_normal_rx_sg_intermed_2}) as follows,
\begin{equation}
	[\hat{\mathbb{L}},  {\bf \hat{x}}^{(k)}] =  \arg\min\limits_{\forall x^{(k)}_i \in\mathcal{G}, \forall {\mathcal{L}} } ||\boldsymbol{y}^n -  \sum_{i=1}^{M_u} \boldsymbol{y}^r_{l_i}x^{(k)}_i ||_{F}^{2} 
	\label{eq:d_mgsm_ml_det}
\end{equation}
where an exhaustive search is performed to find the combination of the reference symbol and the modulated symbol that minimizes (\ref{eq:d_mgsm_ml_det}). The number of bits transmitted per D-MGSM symbol is  
$M_ulog_2{M} + \lfloor log_2{\dbinom{M_t}{M_u}} \rfloor$. 	
The effective channel usage (in \textit{bpcu}) of the differential schemes are added in Table \ref{tab:bpcu_all} to get a better picture about the SE. 
When it comes to the number of transmit antennas required to maintain a given \textit{bpcu} after fixing the modulation order, the proposed schemes fare much better than the conventional differential schemes for SM, as seen in Figure \ref{fig:mt_versus_bpcu}, where it requires only a minimal or sometimes no increase in the number of transmit antennas.

\begin{table}[ht] 	 	
 	\caption{Supported modulation and effective channel usage in terms of bits per channel use (bpcu) for the existing differential schemes along with the proposed schemes }
		\begin{tabular}{l|l l}
		\hline
\multirow{2}{*}{Scheme} & Channel usage  & Modulation          \\	
    	   & 		      (bpcu) 	     & Type      \\
		\hline 
\multirow{2}{*}{DSM \cite{1stdsm}}  &\multirow{2}{*}{$\frac{\lfloor log_2(Mt!) \rfloor + M_tlog_2(M)}{M_t}$}       & \multirow{2}{*}{M-PSK}\\ 
& & \\
 
High rate  &\multirow{2}{*}{$ \frac{\lfloor log_2(M_t!) \rfloor + M_tlog_2(M) + 2M_t}{M_t}$} & \multirow{2}{*}{APSK}   \\
APSK-DSM \cite{dsm_4_ring_apsk}	& &\\	   
GD-SM\cite{gdsm}  & $ \lfloor log_2(M_t) \rfloor + log_2(M)$    & M-QAM\\ 
& & M-PSK\\	
\multirow{2}{*}{Proposed D-GSM}   &\multirow{2}{*}{$\lfloor log_2 \dbinom{M_t}{M_u} \rfloor + log_2(M)$} & M-QAM    \\
& & M-PSK \\
& &\\

\multirow{2}{*}{Proposed D-MGSM}  &\multirow{2}{*}{$\lfloor log_2 \dbinom{M_t}{M_u} \rfloor + M_u log_2(M)$} & M-QAM    \\
& & M-PSK\\
& & \\
		\hline
		\end{tabular}
  		\label{tab:bpcu_all}
 	\end{table} 
 	
\subsection{Power allocation strategy} \label{sec:power_alloc}
The power allocation objective is to transmit the reference symbols at a higher power than the normal symbols and to maximize the average output SNR. It is achieved by allocating equal power to all the $(B-1)$  normal blocks in a frame, and the reference block at higher power, such that the average transmit power  is $\bar{\rho}$ per block. This average power holds for all the modulated symbols of a coherent transmission as well. A frame consists of $M_t$ reference symbols and $K$ information carrying normal symbols totalling to $L = M_t + K$ symbols.

\begin{figure}[h]	
		\begin{center}		
		 \includegraphics[width=8cm, height=2cm]{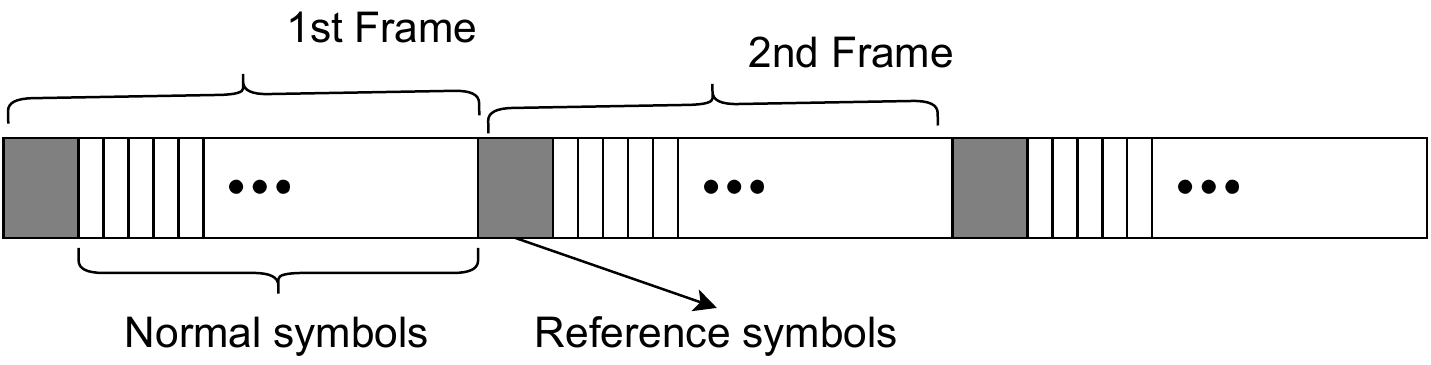} 
    \caption{Frame structure of the proposed schemes }
		  \label{fig:frame_struct}
		\end{center} 
\end{figure}

The average transmit power of the reference block and normal blocks are  $\bar{\rho}_r$ and $\bar{\rho}_n$ respectively, and they are related as, 
\begin{equation}
	\bar{\rho}_r + (B-1)\bar{\rho}_n = B \bar{\rho}
	\label{eq:d_gsm_tot_power_block}
\end{equation} 
where $\bar{\rho}_r > \bar{\rho}_n$.  A frame is divided into one reference block and $(B-1)$ normal blocks. Each of these blocks contains $M_t$ symbols which uses the power allocation strategy similar to differential space-time block code (STBC) in \cite{gd_stbc}. The power allocated for each reference and normal symbol in a block is defined as 
\begin{equation}
\bar{\rho}_{r_s} = \frac{B\bar{\rho}}{(1 + \sqrt{B-1})}
\end{equation}
\begin{equation}
\bar{\rho}_{n_s} = \frac{B\bar{\rho}}{(B-1 + \sqrt{B-1})}.
\end{equation}
Since the proposed scheme requires multiple active antennas to transmit a normal symbol, we define a power allocation scheme that distributes equal power among the $M_u$ active transmit antennas in a symbol from the given power budget of a normal block.
Thus when it comes to each symbol of the normal block, the power is divided equally among the $M_u$ active transmit antennas as
\begin{equation}
\bar{\rho}_{n_{sA}} = \frac{B\bar{\rho}}{(B-1 + \sqrt{B-1}) M_u}
\label{eq:power_per_Mu_ant}
\end{equation}
without disturbing the overall power constraint in (\ref{eq:d_gsm_tot_power_block}). Thus for a given average transmit power of $\bar{\rho}$ per block, the reference and normal blocks follows an SNR described as, $\gamma_r = \frac{1}{\sigma_r^2} = \frac{B\bar{\rho}}{1 + \sqrt{B-1}} $ and $\gamma_n = \frac{1}{\sigma_n^2} = \frac{B\bar{\rho}}{B-1 + \sqrt{B-1}} $ respectivley. Thus for a coherent tranmission the average SNR  during a symbol duration is 
\begin{equation}
	\bar{\gamma} = \bar{\rho} .
\label{eq:equiv_snr}
\end{equation}

\section{Analytical Bit Error Performance} \label{sec:ABEP}
We shall now derive the upper bound of the average bit error probability (ABEP) for both the proposed detectors based on the well known union bounding technique in \cite{simon_digital_2000}. The upper bound
 of the ABEP is given by
\begin{equation} {\mathrm{ABEP}} \le \frac {1}{{m{2^{m}}}}\sum \limits _{{{\mathbf{x}}_{k}} \in \Phi } {\sum \limits _{{{\tilde {\mathbf{x}}}_{k}} \in \Phi } {N({{\tilde {\mathbf{x}}}_{k}},{{\mathbf{x}}_{k}})P({{\mathbf{x}}_{k}} \!\to \! {{\tilde {\mathbf{x}}}_{k}})}} , \end{equation}
where $m$ is the number of bits per symbol. $N(.)$ is the bitwise difference of the symbols ${{\tilde {\mathbf{x}}}_{k}}$ and ${{\mathbf{x}}_{k}}$, and $ \Phi $ is symbol map of D-GSM and D-MGSM schemes. The pairwise error probability (PEP) is $P({{\mathbf{x}}_{k}} \!\to \! {{\tilde {\mathbf{x}}}_{k}})$, which depends on the symbol $\mathbf{x}_{k}$ being transmitted across the channel, and the symbol ${{\tilde {\mathbf{x}}}_{k}}$ is wrongly detected in place of $\mathbf{x}_{k}$. Error occurs when the symbol  ${{\tilde {\mathbf{x}}}_{k}}$ is detected as the best solution  using the minimization expressions of (\ref{eq:d_mgsm_ml_det}) and (\ref{eq:d_gsm_ml_det}), instead of $\mathbf{x}_{k}$.  By using the definition of PEP, we show that, 
\begin{equation} 
 {P}({{\mathbf{x}}_{k}} \to  {{\tilde {\mathbf{x}}}_{k}})= {P}(||{{\mathbf{y}}_{n}} - {{\mathbf{Y}}_{r}}{{\mathbf{x}}_{k}}||_{F}^{2} >\, ||{{\mathbf{y}}_{n}} - {{\mathbf{Y}}_{r}}{{\tilde {\mathbf{x}}}_{k}}||_{F}^{2}) .
 \label{eq:pep_abep_rx_sym_2}
	\end{equation}
When the received symbols ${\bf y}^n = {\bf H}^n {\bf s}^n + {\bf n}^n $ and ${\bf Y}^r = {\bf H}^{r}s^r + {\bf N}^r  $  are substituted in (\ref{eq:pep_abep_rx_sym_2}), we get 
\begin{equation}
 { P({{\mathbf{x}}_{k}} \!\!\to \!\! {{\tilde {\mathbf{x}}}_{k}})} \!=\! {P}(||{\bf n}_n - {\bf N}_r {\bf x}_k||_{F}^{2} \!\!>\!\! ||{\mathbf{H}}{{\mathbf{x}}_{k}} \!-\! {\mathbf{H}}{{\tilde {\mathbf{x}}_{k}}} \!+ \! {\bf n}_n - {\bf N}_r \tilde{{\bf x}_k}||_{F}^{2}). 
\label{eq:pep_abep_rx_sym_3}
\end{equation} 

To further simplify (\ref{eq:pep_abep_rx_sym_3}), we replace the two noise terms by $\mathbf{w}_{1} = {\bf n}_n - {\bf N}_r {\bf x}_k $ and $\mathbf{w}_{2} = {\bf n}_n - {\bf N}_r \tilde{{\bf x}}_k $ and the symbol difference by
$ {\bf d} = \mathbf{H} \mathbf{x_{k}} \!-\! \mathbf{H}\tilde{ \mathbf{x}}_{k} $,  to obtain  

\begin{align}
 { P({{\mathbf{x}}_{k}} \!\!\to \!\! {{\tilde {\mathbf{x}}}_{k}})}   &= {P}( {\bf w}^{H}_{2}{\bf d} + {\bf d}^{H}{\bf w}_2 \!>\! ||{\mathbf{d}}||_{F}^{2}) \notag \\
 &= {P}(\zeta \!>\! ||{\mathbf{d}}||_{F}^{2}), 
 \label{eq:pep_gaussian_form_4} 
 \end{align}
where the norms of ${\bf w}_1$ and ${\bf w}_2$ are statistically the same, and is used one for the other. Now the expression is left with the terms ${\bf w}_2$ and ${\bf d}$ alone. Since the  normalized average power of the symbol is given by $ E\left[ {|{\bf x}_k|}^2 \right]  = 1$, we find that  $\zeta$ is Gaussian distributed as  $\zeta \sim \mathcal{CN}(0, 2(\sigma_n^2 + M_u \sigma_r^2 ))$.  Thus  the conditional PEP (CPEP) in (\ref{eq:pep_gaussian_form_4}) is written in the  Q-function form using $Q(x) = \frac{1}{2\pi}  \int_{x}^{\infty} e^{\frac{-x^{2}}{2}}dx $ as, 
 \begin{equation} { P({{\mathbf{x}}_{k}} \!\!\to \!\! {{\tilde {\mathbf{x}}}_{k}})} \!=\! { {Q\left ({ {\sqrt \frac{||  {{\mathbf{H}}({{\tilde {\mathbf{x}}}_{k}} \!-\! {{\mathbf{x}}_{k}})} ||_{F}^{2}}{ 2(\sigma_n^2 + M_u \sigma_r^2 ) }} }\right )} }.
 \label{eq:pep_Q_fn_form_5}
  \end{equation}
To obtain the actual PEP, we average the CPEP in 
(\ref{eq:pep_Q_fn_form_5}) with respect to the probability density function (PDF) of $\gamma = {||  {{\mathbf{H}}({{\tilde {\mathbf{x}}}_{k}} \!-\! {{\mathbf{x}}_{k}})} ||_{F}^{2}}/{ 2(\sigma_n^2 + M_u \sigma_r^2 )}  $, and by using the finite integration form of Q-function,
 $Q(x) = \frac{1}{\pi}  \int_{0}^{\pi /2} \exp\left( {\frac{-x^{2}}{2\sin^2\theta }}\right) d\theta $, for $(x > 0 )$. The averaging of the CPEP over $\gamma$ using its PDF given by ${P_{\gamma}}\left ({{\gamma} } \right)$,  changes  it to an expression having the moment generating function (MGF), ${M_{\gamma}} \left( . \right)$ given by
 \begin{align} P\left ({{\mathbf{x}}_{k}} \to \tilde{\mathbf{x}}_{k}  \right)=&\frac {1}{\pi }\int _{0}^{\frac {\pi }{2}}\! {\int _{ - \infty }^{ + \infty }\! {\exp \left ({{ \!- \frac {\gamma}{{2{\sin ^{2}}\theta }}} }\right)} } {P_{\gamma}}\left ({{\gamma} }\right)d{\gamma}d\theta \!\!\!  \notag \\ 
 =&\frac {1}{\pi }\int _{0}^{\frac {\pi }{2}} {{M_{\gamma}}} \left ({{ - \frac {1}{{2{\sin ^{2}}\theta }}} }\right)d\theta.
  \label{eq:pep_q_fn_mgf_6}
 \end{align} 
We find that $\gamma$ is a linear combination of chi-square distributed random variables with two degrees of freedom. Thus, based on  the characteristic function of a chi-square distributed random variable given in \cite{turin_1960}, we write the MGF  of a linear combination of chi-square distributed random variables with two degrees of freedom as
\begin{equation}
	{M}_\gamma(s) = \prod^N_{n=1} \frac{1}{1-s \lambda_n}
	\label{eq:mgf_chi_sq}
\end{equation}
where $\lambda_n$'s are the non-zero eigen values of the covariance matrix of $ \frac{ {{\mathbf{H}}({{\tilde {\mathbf{x}}}_{k}} \!-\! {{\mathbf{x}}_{k}})} } {\sqrt{ 2(\sigma_n^2 + M_u \sigma_r^2 ) }}$  and $N = M_r$ is the number of entries in the diagonal of the covariance matrix. All the eigen values are found to be equal and is given as  $ \lambda_1 = \ldots  = \lambda_{M_r} =  \frac{\| {{{\Upsilon }_{v_{1}}}}   \|_{2}^{2} } { 2(\sigma_n^2 + M_u \sigma_r^2 ) } + \cdots +  \frac{\| {{{\Upsilon }_{v_{b}}}}   \|_{2}^{2} } { 2(\sigma_n^2 + M_u \sigma_r^2 ) } + \cdots + \frac{\left \|{ {{{\Upsilon }_{v_{\beta} }}} }\right \|_{2}^{2} } { 2(\sigma_n^2 + M_u \sigma_r^2 ) } $, where $\lambda_i$'s turn out to be the non-zero entries in the diagonal of  the covariance matrix of $ \frac{ {{\mathbf{H}}({{\tilde {\mathbf{x}}}_{k}} \!-\! {{\mathbf{x}}_{k}})} } {\sqrt{ 2(\sigma_n^2 + M_u \sigma_r^2 ) }}$ and the integer value of $\beta$ is in the range of $[2, 2M_u]$. For medium-to-high SNR since $\lambda_n >> 1$, we approximate the MGF of (\ref{eq:mgf_chi_sq}) as 
$ {M}_\gamma(s)\approx \prod^N_{n=1} \left(\frac{1}{-s \lambda_n} \right) $, and the integral of (\ref{eq:pep_q_fn_mgf_6}) is evaluated using  $ \int _{0}^{{\pi \mathord {\left /{ {\vphantom {\pi 2}} }\right . } 2}} {{\sin ^{2m}}} xdx = \frac {(2m - 1)!!}{(2m)!!}\frac {\pi }{2}$ and by substituting $s = - \frac {1}{{2{\sin ^{2}}\theta }}$ to get the final PEP expression as, 
\begin{equation} 
P({{\mathbf{x}_k}} \!\!\to \!\! {{\tilde {\mathbf{x}}}_k}) <   \frac {\left( 2 (\sigma_n^2 + M_u \sigma_r^2) \right)^{{ {M_{r}}}} {{2^{ {M_{r}}}}(2 {M_{r}} - 1)!!}}{{2{{\left(\left \|{ {{{\Upsilon }_{v_{1}}}} }\right \|_{2}^{2} + \cdots +\left \|{ {{{\Upsilon }_{v_{\beta} }}} }\right \|_{2}^{2}\right)}^{M_{r}}}(2 {M_{r}})!!}}.
\label{eq:pep_final_7}
 \end{equation} 

Each of the $M_r$ rows of the covariance matrix of ${\bf d}  = \mathbf{H} \left( \tilde{\mathbf{x}}_{k} - \mathbf{x}_{k} \right)$ is given by $||\Upsilon_{v_{1}}||^2_2 + \cdots + ||\Upsilon_{v_{\beta}}||^2_2 $ and only the non-zero entries of $ {\bf d}$ are considered in (\ref{eq:pep_final_7}). The expression for PEP in (\ref{eq:pep_final_7}) contains the difference of the symbols $\left( {\bf x}_k , \tilde{{\bf x}}_k \right)$ which is either the D-GSM symbols $\left( {\bf x}^{(k)}_{G} \neq  \tilde{{\bf x}}^{(k)}_{G} \right)$  or the D-MGSM  symbols  $\left( {\bf x}^{(k)}_{MG} \neq  \tilde{{\bf x}}^{(k)}_{MG} \right)$ given in Section \ref{sub:d_gsm} and \ref{sub:d_mgsm} respectively.

\section{Computational Complexity analysis} \label{sec:complexity}
We use the total number of real multiplications and additions involved in the detection steps to compare the computational complexity of the proposed schemes with the existing schemes such as GD-SM \cite{gdsm}, and coherent GSMs \cite{gsm1}, \cite{gsm_2012}. These computations are also called floating point operations (flops), which counts all the real additions and  multiplications  required for an operation. For a fair comparison with the above mentioned schemes, we choose to compare the complexity of decoding $K$ information carrying symbols of all the above mentioned schemes. Thus for the proposed   D-GSM scheme, the computations for the different steps of the detector in (\ref{eq:d_gsm_ml_det}) is given by: (\romannum{1}) $M{N}(4M_r + 2M_rM_u)$ operations when ${\bf y}^r_{l_i}\hat{x}^{(k)}$ is performed $M_u$ times and the same is reused all throughout a frame. (\romannum{2}) The subtraction operation of ${\bf y}_n - (.)$ involves $2M{N}M_rK $ operations  and (\romannum{3}) the Frobenius norm involves $4M_r-1$ operations and it is repeated $M{N}K$ times, totalling to $(4M_{r}-1)M{N}K$ flops.
\begin{table}[h]
	\centering	
	\caption{Comparison of percentage change in detection complexity of D-GSM with existing scheme }	   
	\begin{tabular}{|c|c|c|c|}
	 \hline	   
	  Spectral Efficiency  & ($M_t \times M_r$),  & \multicolumn{2}{c|}{Flops w.r.t GSM-1,  \cite{gsm1} } \\ 	  
	   \cline{3-4}	  	   
	   
	 	 (SE) (bpcu)& $M_u = 2$ &	$K =100$ & $K = 400$  \\
		\hline	  
	   ${\bf 5}$& $4 \times 2$ & {$ 3\% \downarrow $}  &   {$ 0.8\% \downarrow $}  \\
	   \hline
	   ${\bf 6}$& $5 \times 2$ &  {$ 2\% \downarrow $}  &  {$ 0.5\% \downarrow $}  \\
	    \hline
	   ${\bf 7}$& $5 \times 2$ &  {$ 0.9\% \downarrow $} &  {$ 0.2\% \downarrow $}  \\  
	    \hline	
	   ${\bf 8}$& $5 \times 2$ & {$ 0.5\% \downarrow $} &   {$ 0.12\% \downarrow $}  \\
  	    \hline	 	 
	  \end{tabular} 	
	\label{tab:complexity_d_gsm}	
\end{table}
 It is seen that there is a marginal decrease in complexity for  D-GSM compared to GSM-1 as shown in Table \ref{tab:complexity_d_gsm}, where the $\downarrow$ denotes the percentage decrease in complexity and  the percentage increase in detection complexity is denoted by $\uparrow$. 
In the case of the proposed D-MGSM scheme, the computations involved for the different steps of the ML detector in (\ref{eq:d_mgsm_ml_det}) are: (\romannum{1}) $2^{M_u}M{N}({8}M_rM_u - 2M_r) $ operations when ${\bf y}^r_{l_i}\hat{x}^{(k)}_i$ is performed $M_u$ times and the same is stored and reused all throughout a frame. (\romannum{2}) The subtraction operation of ${\bf y}_n - (.)$ involves $2^{M_u + 1}M{N}M_rK $ operations  and (\romannum{3}) the Frobenius norm involves $4M_r-1$ operations and it is repeated $2^{M_u}M{N}K$ times, totalling to $(4M_{r}-1)2^{M_u}M{N}K$ flops, where $N = 2^{\lfloor  log_2{M_c} \rfloor} $ and  $M_c = \dbinom{ M_t}{M_u} $. 

\begin{table}[h]
	\centering	
	\caption{Comparison of percentage change in detection complexity of D-MGSM with coherent schemes }	   
	\begin{tabular}{|c|c|c|c|}
	 \hline	   
	  SE& ($M_t \times M_r$),  & \multicolumn{2}{c|}{Flops w.r.t GSM-2,  \cite{gsm_2012} }  \\ 	  
	   \cline{3-4}		  
	   
	 	(bpcu)& $M_u=2$ & $K =100$ & $K = 400$  \\
		\hline	  
	${\bf 5}$ &    $5 \times 2$ & {$ 0.7\% \downarrow $}  &   {$ 0.2\% \downarrow $}  \\
	   \hline
	 ${\bf 6}$&  $4 \times 2$ &  {$ 0.56\% \downarrow $}  &  {$ 0.15\% \downarrow $}  \\
	    \hline
	 ${\bf 7}$&  $5 \times 2$ &  {$ 0.35\% \downarrow $} &  {$ 0.07\% \downarrow $}  \\  
	    \hline	
	 ${\bf 8}$&  $4 \times 2$ & {$ 0.3\% \downarrow $} &   {$ 0.07\% \downarrow $}  \\
  	    \hline	 	 
	  \end{tabular} 	
	\label{tab:complexity_d_mgsm}	
\end{table}

\begin{table}[h]
	\centering	
	\caption{Comparison of percentage change in detection complexity of D-MGSM with GD-SM schemes }	   
	\begin{tabular}{|c|c|c|c|c|}
	 \hline	   
	 SE& GD-SM  \cite{gdsm} & D-MGSM &  \multicolumn{2}{c|}{Flops w.r.t GD-SM} \\
	  \cline{4-5}
	  (bpcu)  & $M_t \times M_r,M$	   &$M_t \times M_r,M$ & $K=100$ &$K=400$ \\
		\hline	  
	   ${\bf 5}$  & $4 \times 2,8$  &    $5 \times 2,2$  & $102\% \uparrow$ & $100\% \uparrow$\\
	   \hline
	    ${\bf 6}$  &  $4 \times 2,16$ &  $4 \times 2,4$  & $1.4\% \uparrow $ & $0.36\% \uparrow$\\
	    \hline
	   ${\bf 7}$  &  $4 \times 2,32$ &   $5 \times 2,4$  & $1.4\% \uparrow$ & $0.36\% \uparrow$\\  
	    \hline	
	    ${\bf 8}$  & $8 \times 2,32$ &   $4 \times 2,8$  & $ 49.3\% \downarrow$ & $ 49.8\% \downarrow$\\
  	    \hline	 	 
  	      ${\bf 9}$  & $8 \times 2,64$ &   $5 \times 2,8$  & $ 49.3\% \downarrow$ & $ 49.8\% \downarrow$\\
  	      \hline
  	      ${\bf 10}$  & $16 \times 2,64$ & $7 \times 2,8$  & $ 49.3\% \downarrow$ & $49.8\% \downarrow$\\
  	      \hline
	  \end{tabular} 	
	\label{tab:complexity_gdsm_d_mgsm}	
\end{table}
The percentage change in flops which denotes  the complexity for the D-MGSM scheme  is added in Table \ref{tab:complexity_d_mgsm} for $M_u = 2$, where the comparison is made with GSM-2. We see a reduction in complexity in comparison to GSM-2, and in Table \ref{tab:complexity_gdsm_d_mgsm} the complexity fares better, since it decreases for higher SE starting  from $8$ bpcu onwards when compared with GD-SM systems. The complexity of the various schemes  using ML based detectors are as follows,
\begin{align}		 	
	 \begin{split} \label{eq:complexity_ml_gd_sm}  
	 		C_{GD-SM} =  {M_t}M \left( 6M_r + 6M_r{K} - K \right)	
	 \end{split} \\	 
	 \begin{split} \label{eq:complexity_ml_gsm1}
	 C_{GSM-1} = M{N} \left( 2M_rM_u + 6M_rK + 4M_r - K \right) + \cdots  \\ + P_l(6M_r+12) + 6M_t 	 \end{split} \\
	 \begin{split} \label{eq:complexity_ml_gsm2}
 C_{GSM-2} = 2^{M_u}M{N} \left(8M_rM_u + 6M_rK - 2M_r -K\right) + \cdots \\ + P_l(6M_r+12) + 6M_t  \end{split} \\
 \begin{split} \label{eq:complexity_ml_d_mgsm}
 C_{D-MGSM} = 2^{M_u}M{N} \left(8M_rM_u + 6M_rK - 2M_r -K\right)
 \end{split} \\
	 \begin{split} \label{eq:complexity_ml_d_gsm}
	 	C_{D-GSM} = M{N} \left( 2M_rM_u + 6M_rK + 4M_r - K \right)
	 \end{split}
\end{align}
where  GSM-1 \cite{gsm1} and  GSM-2 \cite{gsm_2012} are  the coherent counterparts of the proposed schemes  D-GSM and D-MGSM respectively. The complexity of these coherent schemes involves the computations required  for the least-squares (LS) estimation of the channel coefficients as well, where the pilot length is assumed to be $P_l=4M_t$ for all the transmit antennas combined.

\section{Simulation Studies and Discussion }  \label{sec:simulation}

\begin{table}[H]
	\centering	
	\caption{Simulation Parameters}
	\begin{tabular}{l l}	
	\hline	 	
	Channel 		&  Rayleigh fading \\
	Modulation (M) &  ($4, 8, 16, 32, 64$) \\
				   &   MPSK, MQAM  \\	
	Number of transmit antennas ($M_t$) & $4, 5, 6$ \\		
	Number of receive  antennas ($M_r$) & $2$, $3$, $4$ \\
	Number of active  antennas ($M_u$) & $2, 3$ \\
	Information symbols per frame ($K$) & $100$ \\ 
	Pilot length per frame(coherent) ($P_l$) & $4M_t$ \\
	\hline
	\end{tabular} 	
	\label{tab:siml_param}
	\end{table}
	
\begin{figure}[h]	
		\begin{center}		
		  \includegraphics[width=8.5cm, height=8.5cm]{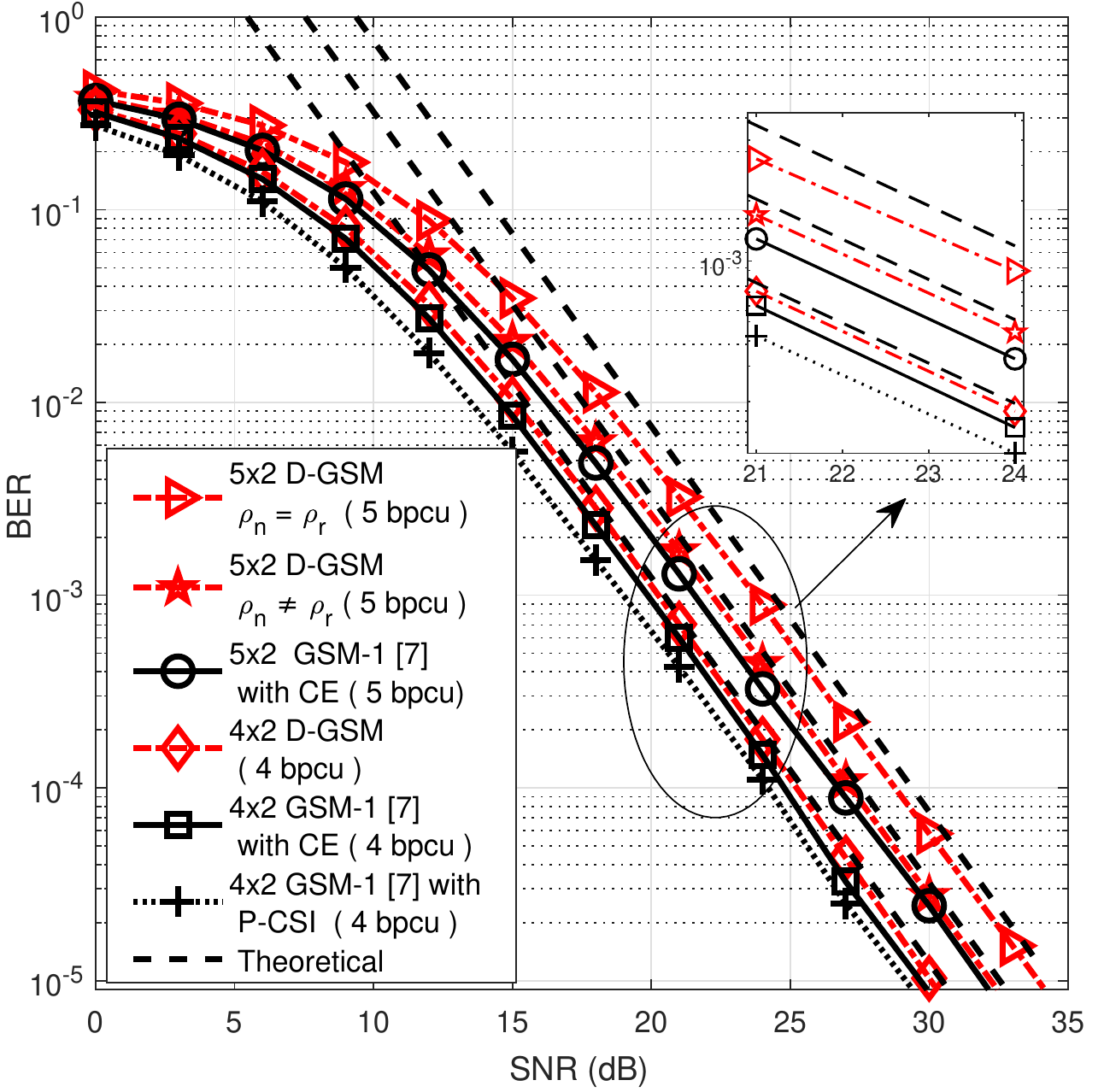}           
		    \caption{Comparison of BER of D-GSM receiver with its equivalent  coherent scheme in a Rayleigh fading channel for  $4$ and $5$ bpcu }
		  \label{fig:d_gsm_gd_sm_gsm1_4bpcu_5bpcu}
		\end{center} 
	\end{figure}
		
In this section, we shall compare the error performance of the proposed schemes D-GSM and D-MGSM with their coherent counterparts \cite{gsm1}, \cite{gsm_2012}, along with the GD-SM \cite{gdsm} scheme which is closer to the proposed ones at the encoding stage. Our results are the output of Monte-Carlo simulations and the average SNR per symbol in a frame remains the same for all the schemes considered here. To ensure a fair comparison with the frame-based encoding of the proposed scheme, a block fading Rayleigh channel is considered and the coherent schemes are simulated by following a similar  frame structure as that of  D-GSM and D-MGSM. For all the investigated cases of simulated and theoretical BER, the plot is made   against the average symbol SNR  per receive antenna given by $\bar{\gamma} =\frac{E_s}{N_0}$ in (\ref{eq:equiv_snr}), where $E_s =  E\left[ {|{\bf s}|}^2 \right]$, is the average power per symbol corresponding to the coherent transmission scheme in Section \ref{sec:sm_model_gsm} and $N_0 = \sigma^2$, is the average noise power at each  receive antenna.

For the coherent schemes, the channel coefficients are assumed to be perfectly known at the receiver in some plots, whereas, in others, it is estimated using LS estimation at the receiver \cite{ofdm_CE_1999, sm_rls_CE_2009}, using the pilots transmitted in the first $P_l$  times slots of every frame.
\begin{table}[h]
	\centering	
	\caption{ Comparison of the effective throughput with the existing coherent schemes}	   
	\begin{tabular}{|c|c|c|c|c|}
   \hline	   
	\multirow{2}{*}{$M_t$} & \multicolumn{2}{|c|}{GSM-1 \cite{gsm1}, GSM-2 \cite{gsm_2012}, $P_l = 4M_t$ }   &  \multicolumn{2}{|c|}{Proposed schemes} \\
	 \cline{2-5}
	 & $K =100$ & $K = 400$ & $K =100$ & $K = 400$ \\
	 \hline
	 $4$   & $86.2\%$  & $96.1\%$  & {$\bf 96.1\%$} & {$\bf 99\%$}     \\
	 \hline
	 $5$  &  $83.3\%$  & $95.2\%$  & {$\bf 95.2\%$} & {$\bf 98.7\%$}     \\
	 \hline
     \end{tabular} 	
	\label{tab:throughput_gsm}	
\end{table}
 Table \ref{tab:throughput_gsm} shows that the proposed differential schemes have a reasonably higher data throughput than the coherent schemes of \cite{gsm1} and \cite{gsm_2012}, where throughput  is defined as,

\begin{equation}
Throughput =  \frac{K} {L}
\label{eq:throughput}
\end{equation}
where the reference signals at the start of a frame are avoided in the numerator. When higher order modulation is considered to achieve higher SE in GD-SM, the number of  transmit antenna is increased which reduces the throughput in comparison to the proposed schemes, and the same is observed in Table \ref{tab:throughput_gd_sm_d_gsms}. Here the  throughput of the proposed schemes are better than GD-SM  when higher order modulation is supported using a larger number of transmit antennas
\begin{figure}[h]	
		\begin{center}		
		  \includegraphics[scale=.65]{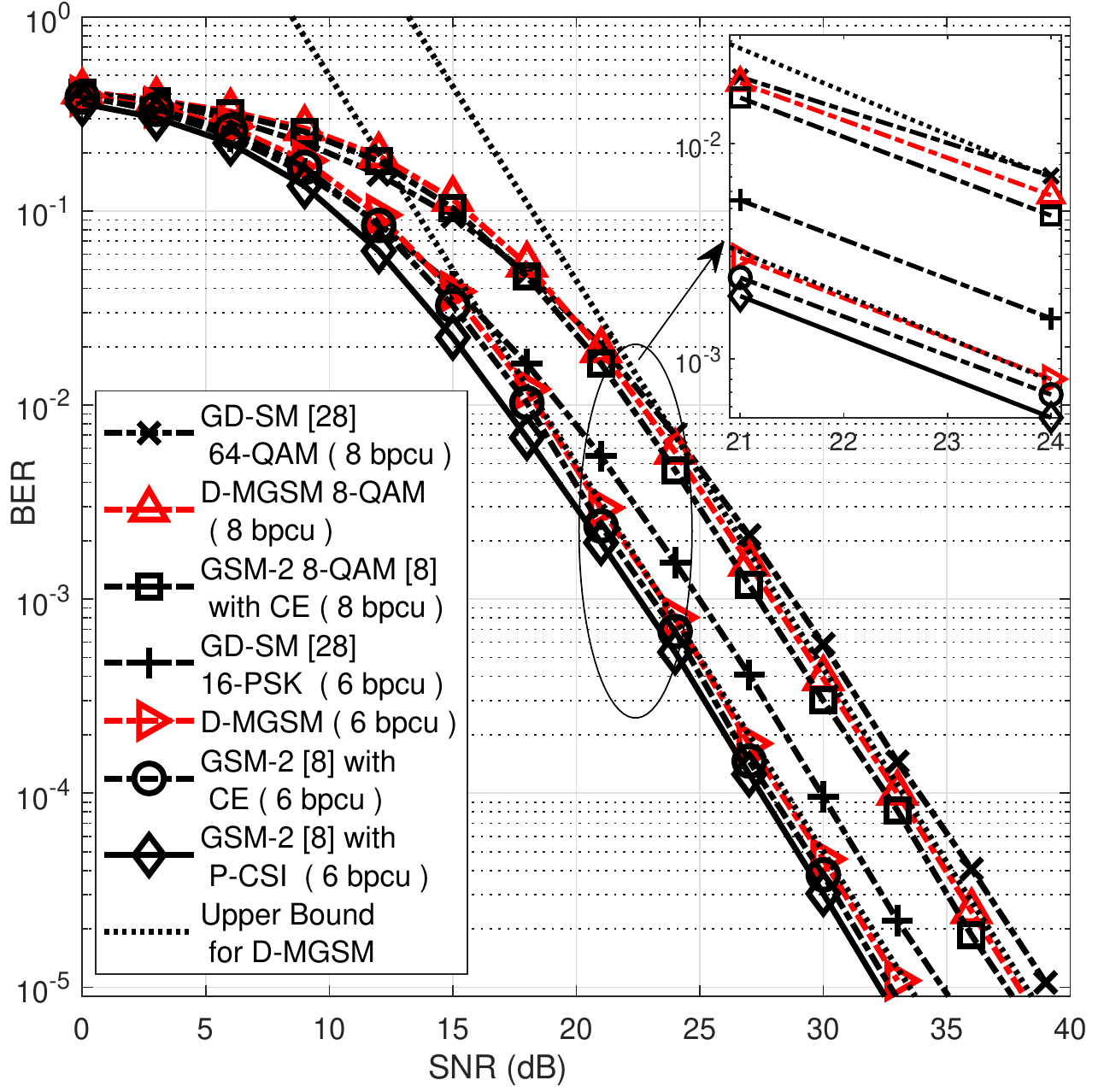}           
		    \caption{Comparison of BER of D-MGSM receiver with the existing detectors for a ($4 \times 2$) system at $6$, $8$ bpcu using MPSK and MQAM }
		  \label{fig:d_mgsm_gd_sm_gsm_2_6_8_bpcu}
		\end{center} 
	\end{figure}

\begin{table}[h]
	\centering	
	\caption{ Comparison of the effective throughput with the GD-SM scheme}	   
	\begin{tabular}{|c|c|c|c|c|c|}
   \hline	   
	\multirow{2}{*}{$M_t$} & \multicolumn{2}{|c|}{GD-SM \cite{gdsm}}   &  \multicolumn{3}{|c|}{Proposed schemes} \\
	 \cline{2-6}
	 &$K =100$ & $K = 400$ & $M_t,M_u$ &  $K =100$ & $K = 400$ \\
	 \hline
	 $4$   & $96.1\%$  & $99\%$ & $4,2$ &{$96.1\%$} & {$99\%$}     \\
	 \hline
	$8$  &  $92.5\%$  & $98\%$  & $5,2$ & {$\bf 95.2\%$} & {$\bf 98.7\%$}     \\	
	 \hline
	 $16$  &  $86.2\%$  & $96.1\%$ & $6,2$ &{$\bf 94.3\%$} & {$\bf 98.5\%$}     \\
	 \hline
     \end{tabular} 	
	\label{tab:throughput_gd_sm_d_gsms}	
\end{table}

\begin{figure}[h]	
		\begin{center}			
		  \includegraphics[scale=.7]{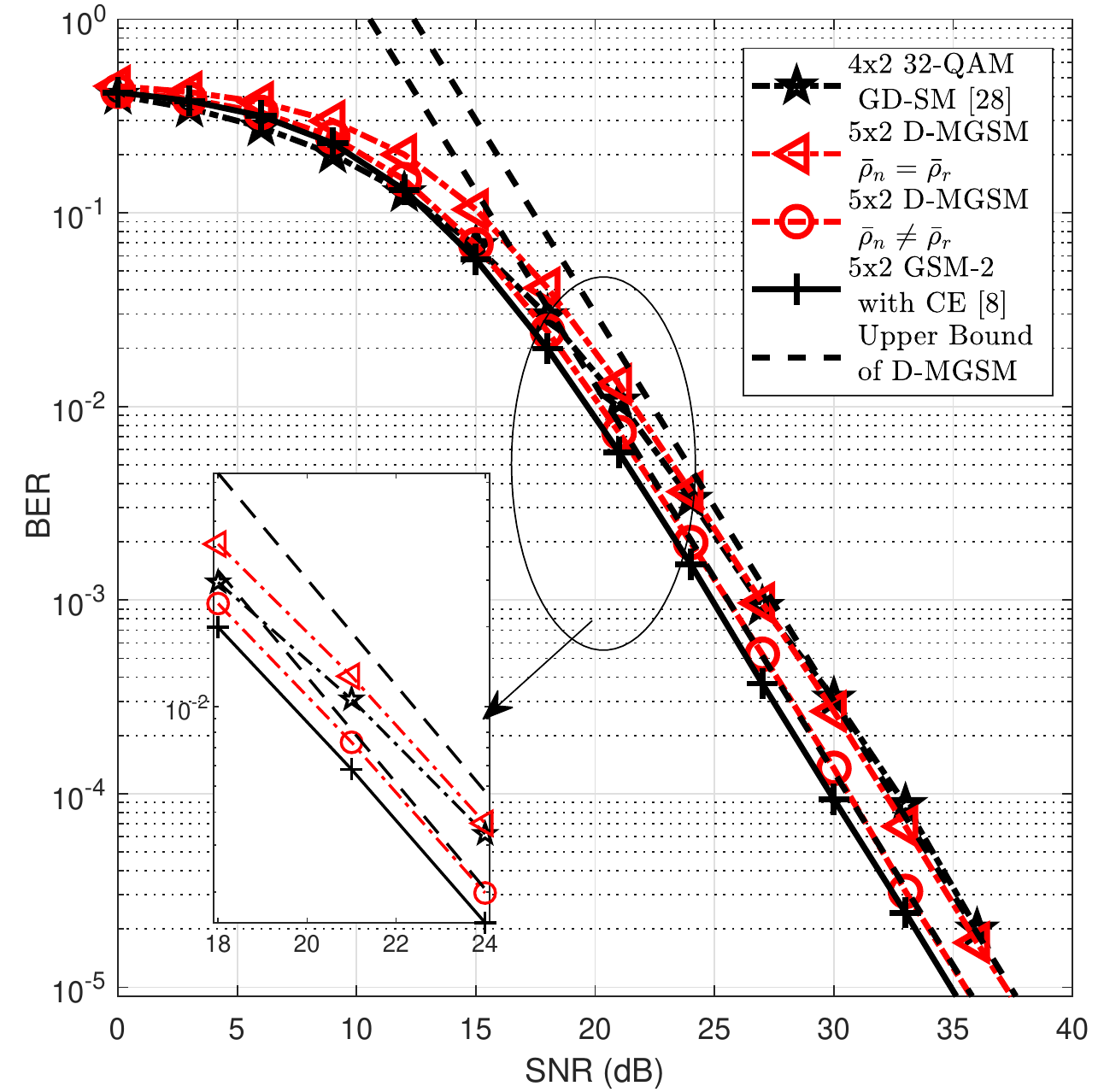}           
		    \caption{Comparison of BER of D-MGSM receiver with the existing detectors for  $7$ bpcu }		    
		  \label{fig:d_mgsm_gd_sm_gsm_2_7_bpcu}		  
		\end{center} 
	\end{figure}
		 	 
\par
 Figure \ref{fig:d_gsm_gd_sm_gsm1_4bpcu_5bpcu} shows the BER  performance of GSM-1 \cite{gsm1}  and the proposed D-GSM for $4 \times 2$ and $5 \times 2$ configurations. It is seen that  the performance penalty of the proposed D-GSM scheme over its coherent  counterpart is less than $0.5$ dB and the upper bound on the ABEP is tight from the mid-SNR region onwards and for the BER of $10^{-2}$ or lesser. When perfect-CSI (P-CSI) is considered at the receiver, the marginal gain in SNR is less than $1$ dB for the coherent scheme. The improvement due to the power allocation strategy ($ \bar{\rho}_r \neq \bar{\rho}_n $) is observed for the $5 \times 2$ system, where the plots without power allocation ($ \bar{\rho}_r = \bar{\rho}_n $) has a penalty of more than $2$ dB with the coherent scheme in Figure \ref{fig:d_gsm_gd_sm_gsm1_4bpcu_5bpcu}.\\
 \par
 In Figure \ref{fig:d_mgsm_gd_sm_gsm_2_6_8_bpcu}, D-MGSM using $8$-QAM is compared with GD-SM employing $64$-QAM for the same SE, and the former has an SNR gain of close to $1$ dB. But when the modulation type of GD-SM is changed to $16$-PSK, the SNR gain of D-MGSM increases to $2$ dB in the case of $6$ bpcu. GSM-2 \cite{gsm_2012} has a marginal gain in SNR in comparison with D-MGSM and it is consistent with that of the gain for the D-GSM scheme in Figure \ref{fig:d_gsm_gd_sm_gsm1_4bpcu_5bpcu}, for both the cases of P-CSI and CE using LS with a pilot length of $P_l =4M_t$. In Figure \ref{fig:d_mgsm_gd_sm_gsm_2_7_bpcu}, the impact of unequal power allocation ($ \bar{\rho}_r \neq \bar{\rho}_n $) is verified for D-MGSM and it has a performance gain of $1$ dB in comparison with the case of equal power allocation ($ \bar{\rho}_r = \bar{\rho}_n $) for the $5 \times 2$ system. The upper bounds of the BER are plotted for both types of power allocation, and the gain with the GD-SM scheme using $32$-QAM is close to $1$ dB when unequal power ($ \bar{\rho}_r \neq \bar{\rho}_n $) is employed. The GSM-2 scheme using LS based CE has a gain of just $0.5$ dB with the D-MGSM scheme. Generally, it is observed that the penalty in error performance with the coherent scheme is negligible and the proposed scheme fares better than GD-SM by $1$ to $2$ dB for different modulation types. Since certain approximations are made in the derivation of ABEP for medium-to-high SNR, the bounds become tighter from the medium SNR region onwards. 
 
 \begin{figure}[h]	
		  \includegraphics[scale=.8]{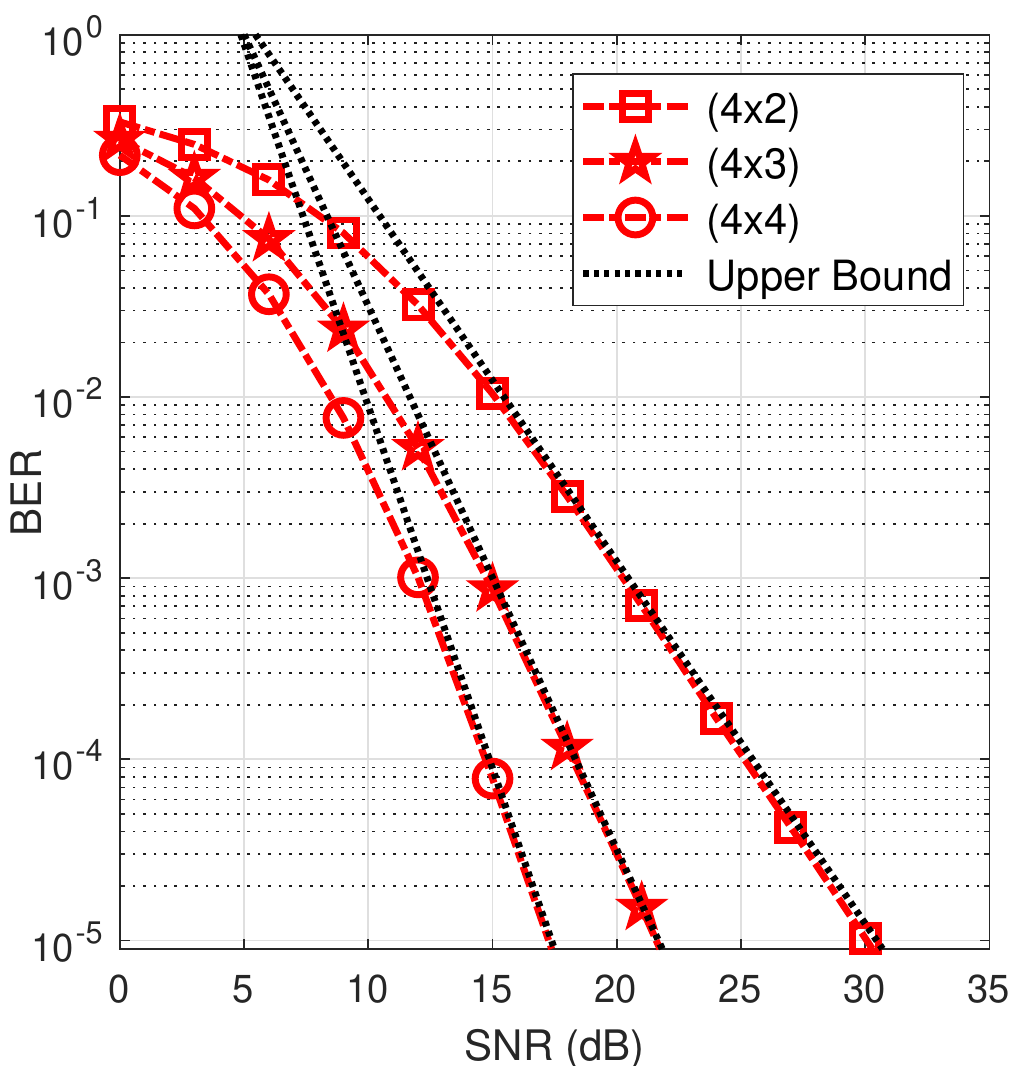}           
		    \caption{BER of the proposed D-GSM receiver with $M_r$ = $2$, $3$, $4$ for $4$ bpcu }
		  \label{fig:rx_div_d_gsm}
	\end{figure}
		
\begin{figure}[h]	
		\begin{center}		
		  \includegraphics[scale=.8]{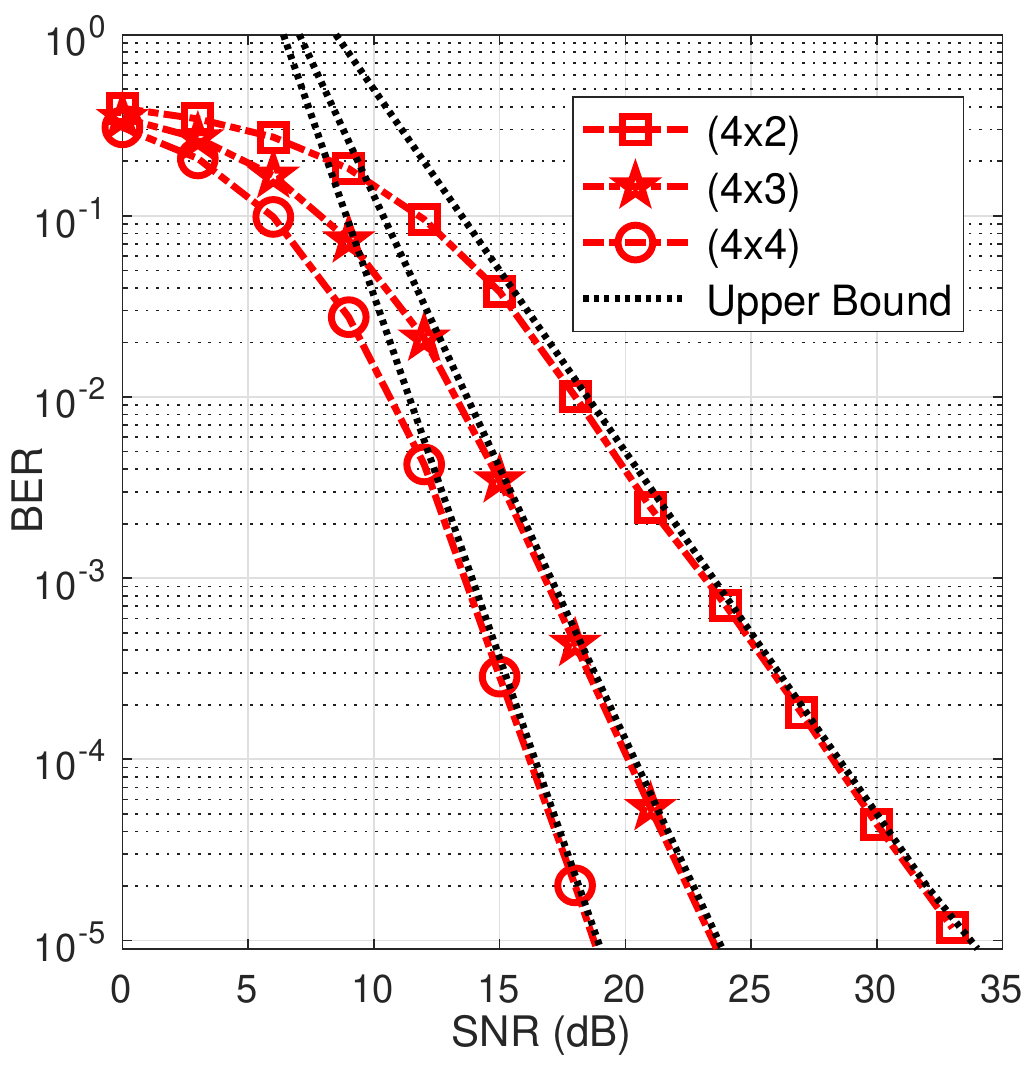}           
		    \caption{BER of the proposed D-MGSM receiver with $M_r$ = $2$, $3$, $4$ for $6$ bpcu}
		  \label{fig:rx_div_d_mgsm}
		\end{center} 
	\end{figure}

\begin{figure}[h]	
		  \includegraphics[scale=.7,center]{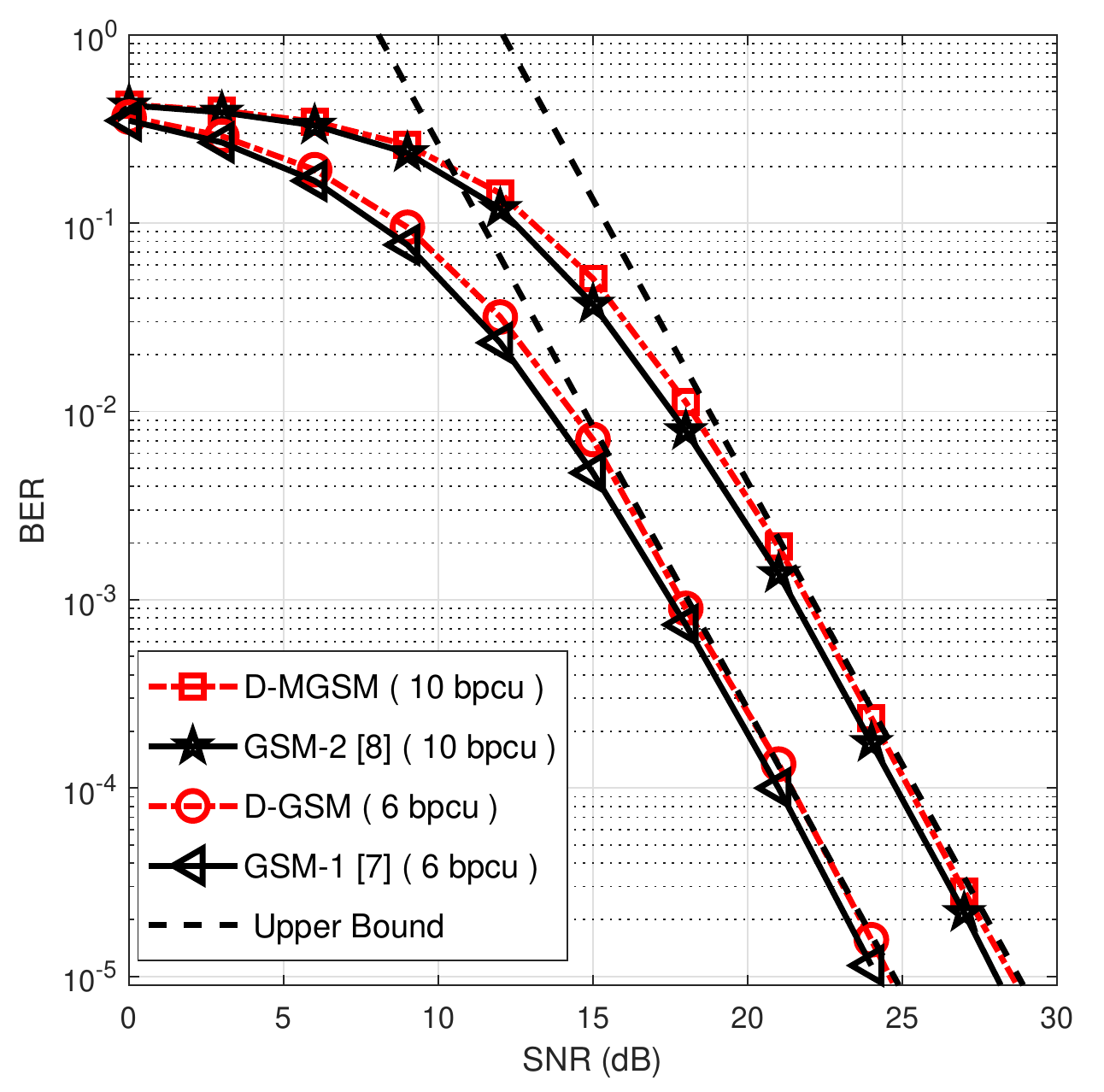}           
		    \caption{BER of  coherent and differential detectors  using a $6 \times 3$ system with $M_u =3$ active antennas}
		  \label{fig:high_tx_ant_gsms}
	\end{figure}

We also verify the derived upper bound for both the proposed schemes and find from Figure \ref{fig:rx_div_d_gsm} and \ref{fig:rx_div_d_mgsm} that the simulated BER graphs closely achieve the bound as the receiver antennas are increased as $M_r = 2$, $3$, $4$. The improvement in error probability is reflected in the derived bound of the PEP in (\ref{eq:pep_final_7}), where the  term $\frac {\left( 4 (\sigma_n^2 + M_u \sigma_r^2) \right)^{{ {M_{r}}}} }{{{{(\left \|{ {{{\Upsilon }_{v_{1}}}} }\right \|_{2}^{2} + \cdots +\left \|{ {{{\Upsilon }_{v_{\beta} }}} }\right \|_{2}^{2})}^{M_{r}}}}}$ from (\ref{eq:pep_final_7}) has a form similar to $\frac{1}{(SNR)^{M_r}}$, and thus offers a diversity order of $M_r$. Thus for a given SNR in the medium-to-high range, the PEP value decreases for an increase in $M_r$, thereby lowering the BER relative to the number of receive antennas. The proposed schemes demonstrate consistency in error performance compared to the coherent schemes, even for higher number of transmit ($M_t$) as well as active antennas ($M_u$), and the same is observed in Figure  \ref{fig:high_tx_ant_gsms}, where the identical symbols in the active antennas of D-GSM  improves the SNR gain by $4$ dB in comparison to the D-MGSM scheme for the same antenna configuration and modulation order. This improvement owing to the number of active antennas is explained using the analytical PEP expression of (\ref{eq:pep_final_7}), where the average value of the parameter $\beta \in [2, 2M_u]$, is close to the upper bound of $2M_u$ for D-GSM, thus ensuring a smaller PEP value in comparison to the D-MGSM scheme where $\beta$ is closer to the lower bound of $2$. The denominator term of (\ref{eq:pep_final_7}) given by ${\left(\left \|{ {{{\Upsilon }_{v_{1}}}} }\right \|_{2}^{2} + \left \|{ {{{\Upsilon }_{v_{2}}}} }\right \|_{2}^{2} + \cdots +\left \|{ {{{\Upsilon }_{v_{\beta} }}} }\right \|_{2}^{2}\right)^{M_r}}$, is influenced by the value of $\beta$, which depends on the encoding signal in the  active antennas corresponding to the two schemes.
			
The proposed schemes are a convenient alternative to the coherent schemes as seen from the marginal penalty in BER. Also, the schemes have a minor reduction in detection complexity and a nominal increase in the effective data throughput in comparison with the coherent schemes. Even though the detection complexity concerning GD-SM seems higher at lower SE, our schemes fare better when higher SE is considered. Moreover, they achieve significant improvement in BER over GD-SM schemes.

\section{Conclusion}  \label{sec:conclusion}
In this paper, we have proposed two differential schemes for SM which utilize multiple active antennas to increase the transmission rate. Furthermore, the analysis of the asymptotic ABEP is also performed based on which a suitable antenna configuration can be chosen conveniently in a real-world deployment. Since these
 schemes dispense with the knowledge of CSI at the receiver, the processing power and the effective bandwidth are also conserved. Also, the new differential schemes have an error performance close to the corresponding coherent schemes and they fare better in their characteristics than most of the  conventional single active antenna based differential schemes. In the context of devices supporting higher SE, there is a need for low complexity detection algorithms, as the optimal decoders are computationally intensive for higher order modulation. There is also a need for improving the error performance at lower SNRs and bridging the marginal penalty in SNR that is presently seen in comparison to the single antenna based differential modulation. One possible solution is to  modify the order or combination of the transmit antennas in the reference block to reduce the ambiguity of the TAC. But, in general, our schemes support user equipments having a limited form factor, without compromising on higher SE requirements, by utilizing the limited number of transmit antennas sharing the available  RF chains. Also, the new schemes have a higher data rate and slightly lower detection complexity in comparison with the coherent schemes. Thus, with the increasing demand for cost-effectiveness and power efficiency for futuristic communication devices, the schemes proposed in this paper are very relevant in the context of green communication systems which need to support higher data rates with lesser computational complexity.

\section{Acknowledgement}
The authors used the Advanced Wireless Communication \& Signal Processing lab, ECE, NIT Calicut created under the scheme \lq{Fund for Improvement of Science and Technology}\rq ( SR/FST/ET-I/2017/68) of the Department of Science and Technology (DST), Govt. of India.


\end{document}